\documentclass[final,5p,times,twocolumn]{elsarticle}

\usepackage{amssymb}
\usepackage{amsmath}
\usepackage{graphicx}
\usepackage{float}
\newcommand{\be}{\begin{equation}}
\newcommand{\ee}{\end{equation}}

\journal{Physics Letters B}

\begin{document}

\begin{frontmatter}

\title{Experimental constraints on the second clock effect}


\author{I. P. Lobo}
\ead{iarley\_lobo@fisica.ufpb.br}
\author{C. Romero\corref{cor1}\fnref{label2}}
\ead{cromero@fisica.ufpb.br}
\fntext[label2]{Corresponding author}

\address{Departamento de F\'{i}sica, Universidade Federal da
Para\'{i}ba, Caixa Postal 5008, CEP 58051-970, Jo\~{a}o Pessoa,
PB, Brazil}

\begin{abstract}
We set observational constraints on the second clock effect, predicted by Weyl
unified field theory, by investigating recent data on the dilated lifetime of
muons accelerated by a magnetic field. These data were obtained in an
experiment carried out in CERN aiming at measuring the anomalous magnetic
moment of the muon. In our analysis we employ the definition of invariant
proper time proposed by V. Perlick, which seems to be the appropriate notion
to be worked out in the context of Weyl space-time.

\end{abstract}


\end{frontmatter}

\section{Introduction}\label{intro}
Since the advent of the special and general relativity
the quest for the determination of the true geometric nature of space-time has
long been a debated matter of research among theoretical physicists. The
treatment of space-time as a differential manifold endowed with a Riemannian
metric tensor, which obeys Einstein's field equations, still remains the
paradigm of gravity theory. However, in recent years a great deal of effort
has gone into the investigation of the so-called \textit{modified gravity
theories}, mainly motivated by attempts at explaining current data coming from
observational cosmology as well as the important issues of dark matter and
dark energy \cite{Clifton}. In this letter, however, we revisit some ideas
developed by H. Weyl in his unified theory, one of the first modified gravity
theories, which appeared soon after the birth of general relativity
\cite{weyl1}. Weyl's theory encountered a severe objection put forward by
Einstein, who believed that it would lead to a physical effect not yet
observed (the so-called \textit{second clock effect}). Curiously, as far as we
know, neither theoretical calculations nor any experimental attempt at
measuring the magnitude of the predicted effect has been carried out up to
now.  

Let us now briefly recall some basic tenets of the geometry conceived by H.
Weyl which underlies his unified theory. Perhaps the main feature of this
geometry is the fact that a vector can have its length changed when parallel
transported along a curve, which is a consequence of the presence of a 1-form
field in the compatibility condition between the metric and the affine
connection. The existence of a group of transformations that leaves this new
compatibility condition invariant is another interesting fact noticed by Weyl,
which ultimately led to the discovery of the gauge theories
\cite{ORaifeartaigh:1998ddh}. As is well known, Weyl's idea was to give a
geometric character to the electromagnetic potential by identifying it with a
purely geometric 1-form field. He then proposed an invariant action that
contained both the gravitational and the electromagnetic fields. However,
Einstein pointed out that the non-integrability of length, a characteristic of
Weyl space-time, would imply that the rate at which a clock measures time,
i.e. its clock rate, would depend on the past history of the clock. As a
consequence, spectral lines with sharp frequencies would not appear
\cite{weyl1}. This came to be known in the literature as the \textit{second
clock effect} \cite{Brown}. (The \textit{first clock effect} refers to the
well-known effect corresponding to the \textquotedblleft twin paradox"
predicted by special and general relativity theories)

Despite the fact that this essentially qualitative objection has led to a
rejection of Weyl theory as being non-physical, an actual measurement of the
magnitude of the second-clock effect predicted by Weyl theory has never been
carried out. Moreover, worse than that, as far as we know even the concept of
proper time measured by an ideal clock in Weyl theory has never been
discussed, neither by Einstein nor by Weyl himself. In fact, the usual
definition of proper time adopted in general relativity as the arc-length of a
curve (the clock hypothesis) cannot be properly carried over to Weyl geometry
for the simple reason that this definition is not invariant under Weyl
transformations (see \cite{Romero:2015ana,Barcelo:2017tes} and references
therein). It turns out, however, that this problem has been finally settled by
V. Perlick, who proposed a definition of proper time which is consistent with
Weyl's principle of invariance \cite{perlick,avalos}. Perlick's notion of
proper time provides a correction to the arc-length formula, and reduces to
the general relativistic proper time when the Weyl 1-form field vanishes.
Moreover, it can be used to set experimental bounds on the predicted second
clock effect. Following a renewed interest in Weyl theory, we believe that
attempts to detect the possible existence of the second clock effect is of
interest in its own, and may lead to results of physical relevance whose
significance may lie beyond any particular gravity theory.

In this letter, we propose to use as our standard clocks unstable particles by
investigating the effect of an external magnetic field on their dilated
lifetime. Specifically, our aim is to set an experimental constraint on the
second clock effect by looking at the Perlick's proper time corresponding to
the dilated lifetime of muons accelerated by this magnetic field.



\section{Weyl geometry}\label{sec-weyl}

As we have mentioned before, the basic idea of Weyl
geometry is the introduction of a 1-form field $\sigma_{\alpha}$ (called the
Weyl field), which is used to replace the Riemannian compatibility condition
between the metric $g_{\mu\nu}$ and the connection $\nabla_{\alpha}$ by
requiring that the new condition reads
\begin{equation}
\nabla_{\alpha}g_{\mu\nu}=\sigma_{\alpha}g_{\mu\nu}. \label{comp}%
\end{equation}
Weyl then found out that by performing the simultaneous transformations
\begin{subequations}
\label{w-trans}%
\begin{equation}
\bar{g}_{\mu\nu}=e^{f}g_{\mu\nu}, \label{conformal}%
\end{equation}%
\end{subequations}
\begin{equation}
\bar{\sigma}_{\alpha}=\sigma_{\alpha}+\partial_{\mu}f, \label{length}%
\end{equation}
where $f=f(x)$ is an arbitrary scalar function, the compatibility condition
(\ref{comp}) is preserved, i.e., we have $\nabla_{\alpha}\bar{g}_{\mu\nu}%
=\bar{\sigma}_{\alpha}\bar{g}_{\mu\nu}$. The discovery of this invariance is
generally considered to be the birth of modern gauge theories (see
\cite{ORaifeartaigh:1998ddh} and references therein). It turns out then that
the condition (\ref{comp}) leads to a new kind of curvature, given by
$F_{\mu\nu}=\partial_{\mu}\sigma_{\nu}-\partial_{\nu}\sigma_{\mu}$, called by
Weyl \textit{the length curvature}, which is invariant under (\ref{length}).
These findings led Weyl to identify the 1-form $\sigma_{\alpha}$ with the
4-potential $A_{\alpha}$ of the electromagnetic field \cite{weyl1} by writing%

\begin{equation}
\sigma_{\alpha}=\lambda\,A_{\alpha}, \label{lambda1}%
\end{equation}
where the constant $\lambda$ is introduced just for dimensional reasons since
$\sigma_{\alpha}$ has dimensions of $[\text{length}]^{-1}$( of course, it is
always possible to choose units such that $\lambda=1$).

The length curvature can be viewed as a measure of the non-integrability of
vector lengths when a vector\ field is parallel transported around a loop. For
instance, let $V^{\mu}$ be the components with respect to a coordinate basis
of a time-like vector $V$ that is parallel transported around a closed curve
$\gamma\mapsto\gamma(t):\gamma\lbrack a,b]\in R\rightarrow M$ (with
$\gamma(a)=\gamma(b)$). If we denote $L^{2}=g_{\mu\nu}V^{\mu}V^{\nu}$ then it
can easily be shown that
\begin{equation}
L(a)=L(b)\exp\left[  \frac{1}{2}\oint\sigma_{\mu}\frac{dx^{\mu}}{dt}dt\right]
,
\end{equation}
\ where $d\gamma/dt\doteq(dx^{\mu}/dt)\partial_{\mu}$, \ $L(a)$ and $L(b)$
denote the initial and final length of $V$, respectively. Surely, $L(a)=L(b)$
if and only if there exists a scalar function $\phi$ such that $\sigma_{\mu
}=\partial_{\mu}\phi$. Clearly, in this case, from Stokes' theorem, $F_{\mu
\nu}=\partial_{\mu}\sigma_{\nu}-\partial_{\nu}\sigma_{\mu}$ must vanish and we
end up with a \textit{Weyl Integrable Space-Time (WIST)}.

We could say that the non-integrability of lengths is in the root of the
already mentioned Einstein's objection to Weyl's theory. Indeed, Einstein
argued that this predicted effect implies that the clock rate of atomic clocks
should be path dependent. In fact, Einstein's reasoning is based on two hypothesis:

a) The proper time $\Delta\tau$ measured by a clock travelling along a curve
$\gamma=\gamma(t)$ is given as in general relativity, that is, by the
(Riemannian) prescription
\begin{equation}
\Delta\tau=\frac{1}{c}\int\left[  g(V,V)\right]  ^{\frac{1}{2}}dt=\frac{1}%
{c}\int\left[  g_{\mu\nu}V^{\mu}V^{\nu}\right]  ^{\frac{1}{2}}dt,
\label{proper time}%
\end{equation}
where $V$ denotes the vector tangent to the clock's world line and $c$ is the
speed of light. This assumption is known as the \textit{clock hypothesis }and
assumes that the proper time only depends on the instantaneous speed of the
clock and \ on the metric field.

b) The fundamental clock rate of standard clocks is given by the (Riemannian)
length $L=$ $\sqrt{g(V,V)}$ of a certain vector $V$.

However, it has been argued recently that in order to discuss the existence of
the second clock effect a new notion of proper time, consistent with\textit{
Weyl's Principle of Gauge Invariance }\footnote{The \textit{Principle of Gauge
Invariance} asserts that all physical quantities must be invariant under the
gauge transformations. This principle was strictly followed by Weyl and guided
him\ when he had to choose an action for his theory. It should also be noted
here that any invariant scalar of this geometry must necessarily be formed by
both the metric $g_{\mu\nu}$ and the Weyl gauge field $\sigma_{\mu}$.}\textit{
,} is needed \cite{Romero:2015ana}. It happens to be that such a notion exists
and was recently given by V. Perlick \cite{perlick}. \ 

Let us now briefly recall the notion of proper time proposed by V. Perlick.
First, let us define a \textit{standard clock }according to the following
definition: A time-like curve $\gamma:\gamma\lbrack a,b]\in R\rightarrow M,$
$t\mapsto\gamma(t)$, is called a \textit{standard clock} if $\frac
{D\gamma^{\prime}}{dt}$ is orthogonal to $\gamma^{\prime}(t)$, i.e.
$g(\gamma^{\prime},\frac{D\gamma^{\prime}}{dt})=0$. We will then say that a
time-like curve $\gamma$ is parametrized by proper time if the parametrized
curve is a standard clock. It can be shown that from this definition it
follows that the proper time elapsed between two events corresponding to the
parameter values $t_{0}$ and $t$ in the curve $\gamma$\ is given by

\begin{align}
&  \Delta\tau(t)=\label{propertime1}\\
&  \left(  \frac{d\tau/dt}{\sqrt{g_{\alpha\beta}\dot{x}^{\alpha}\dot{x}%
^{\beta}}}\right)  _{t=t_{0}}\int_{t_{0}}^{t}\exp{\left(  -\frac{1}{2}%
\int_{u_{0}}^{u}\sigma_{\rho}\dot{x}^{\rho}ds\right)  }\left[  g_{\mu\nu}%
\dot{x}^{\mu}\dot{x}^{\nu}\right]  ^{1/2}du,\nonumber
\end{align}
where the overdot means derivative with respect to the curve's parameter
\cite{avalos} . It has also been shown that Perlick's time has all the
properties a good definition of proper time in a Weyl space-time should have,
such as, Weyl-invariance, positive definiteness, additivity. In addition to
that, in the limit in which the length curvature $F_{\mu\nu}$ goes to zero
Perlick's time\ reduces to the Riemannian or WIST proper time. Recently, it
was shown the equivalence between this definition and the one given in the
well-known paper by Ehlers, Pirani, and Schild (EPS) \cite{avalos,eps}. The
latter was entirely based on axiomatic approach which leads to a Weyl
structure as the most suitable model for space-time.

\begin{figure}[ptb]
\hfill\includegraphics[scale=0.2]{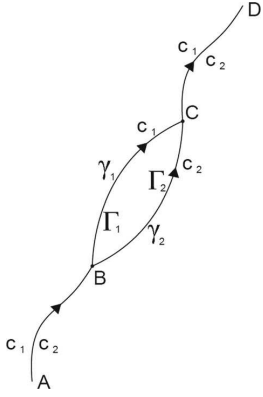}\hspace*{\fill}%
\caption{Synchronized clocks $c_{1}$ and $c_{2}$ follow world lines
$\gamma_{1}$ and $\gamma_{2}$, which are coincident from point $A$ to $B$,
where they are separated to follow the parts $\Gamma_{1}$ and $\Gamma_{2}$ of
these lines until point $C$, where they are once again joined and continue
together until point $D$.}%
\label{fig1}%
\end{figure}

Another important property of Perlick's hypothesis (perhaps unexpected)
concerning the proper time of a standard clock is that it also predicts the
existence of the second clock effect, namely, that the clock rate of a local
observer depends on its path \cite{avalos}. More precisely, consider two
clocks $c_{1}$ and $c_{2}$ synchronized at point $A$ (see Fig.(\ref{fig1})),
which are transported together until point $B$, then separated and transported
along two different paths, $\Gamma_{1}$ and $\Gamma_{2}$, until point $C$,
where they are joined again. These clocks of course will be desynchronized,
i.e., they will measure different times when compared at $C$, and this
constitutes \textit{the} \textit{first clock effect}. However, due to the Weyl
field the clock will not tick at identical rates even when being at rest with
respect to each other at $C$. More precisely, let $\tau_{1,2}$ ,respectively,
be the proper times of clocks $c_{1}$ and $c_{2}$ after they have met at point
$C$. Assuming Perlick's proper time hypothesis, it can be shown that
\cite{avalos}
\begin{equation}
\tau_{1}=\tau_{2}\exp\left(  \frac{1}{2}\int_{\Gamma_{1}}\sigma_{\mu}dx^{\mu
}-\frac{1}{2}\int_{\Gamma_{2}}\sigma_{\mu}dx^{\mu}\right)  .
\end{equation}
This clock-rate discrepancy has been referred to in the literature as
\textit{the} \textit{second clock effect }\cite{Brown}.

It turns out that at the present status of our knowledge we still do not know
whether the second clock effect, a theoretical consequence of Weyl theory,
firstly pointed out by Einstein, does exist. Supposing its existence as a real
physical phenomenon not yet investigated, it seems rather timely to start an
experimental research program to detect\ it. As a first step in this
direction, we have analyzed and used recent data on dilated lifetime of muons
accelerated by a magnetic field as a possible way to set observational
constraints on the second clock effect. Surely, an arena for setting bounds on
this effect requires two elements: i) we should obviously have a controlled
\textquotedblleft clock", or at least a periodic or a time-limited phenomenon
accurately measured; ii) the clock must be placed in a region where there
exists an electromagnetic field. With these in mind, it seems appropriate here
to consider the lifetime of unstable particles, a very well-known phenomenon
which has been measured with high precision.

In this letter, we will consider the particular case of muons accelerated by
an external magnetic field. As we would expect, they will have their lifetime
dilated due to relativistic effects (a manifestation of the first clock
effect). However, by taking into consideration that the proper time of the
muons will now be given by (\ref{propertime1}), we also allow for possible
contributions to the lifetime dilation originated from the second clock
effect. From this analysis we will be able to set the first experimental upper
limit on the value of the Weyl parameter $\lambda$ of Eq.(\ref{lambda1}),
which amounts, in fact, to establish an upper constraint to the existence of
the second clock effect.


\section{Bound from the dilated muon lifetime}\label{sec-muon}
\begin{figure}[ptb]
\hfill\includegraphics[scale=0.3]{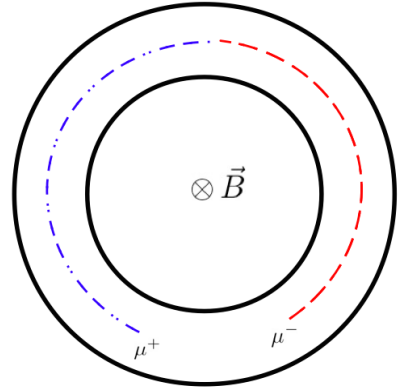}\hspace*{\fill}\caption{For a
magnetic field $\vec{B}$ into the page, the positive muon $\mu^{+}$ follows
the circle in blue (dot-dashed line), while the negative muon follows the red
circle (dashed line). }%
\label{fig2}%
\end{figure}

In the Muon Storage Ring at
CERN the anomalous magnetic moment of positive and negative muons were
measured, whose final report can be found in \cite{Bailey:1978mn}. The
experiment analyzed the orbital and spin motion of highly polarized muons in a
magnetic storage ring, and is described as follows. Protons from a synchrotron
accelerator hit a target and produce pions, which in turn decay into muons
plus a neutrino; the muons are injected into a region with uniform magnetic
field, where they are accelerated to travel in circles (see Fig.(\ref{fig2}))
and have their lifetimes dilated\footnote{There is a quadrupole electrostatic
field that keeps the muon beam aligned, but it does not contribute to our
calculation.}. 

We remark that the same device was also used in the recent experiment E821,
carried out at the Brookhaven National Laboratory (whose final report can be
found in \cite{Bennett:2006fi}), which improved the precision of the CERN's
experiment. However, in addition to measuring the anomalous magnetic moment,
the latter also focused on measuring the first clock effect through the
dilation of the muon's lifetime (which is not the case of E821, whose main
idea was the measurement of the muon's magnetic moment). For this reason we
have chosen the CERN experiment as more appropriate for setting constraints on
the second clock effect.

In what follows we will use Eq.(\ref{propertime1}) together with
(\ref{lambda1}). As a first approximation, we will neglect the gravitational
field of the Earth, and\ hence consider Minkowski space-time metric, written
in cylindrical coordinates $\{t,\rho,\varphi,z\}$. The circular paths followed
by the muons will be parametrized by the laboratory coordinate time $t$, while
the initial conditions, the electromagnetic potential and the velocity will be
chosen as
\begin{subequations}
\begin{align}
t_{0}  &  =0,\\
\left( \frac{d\tau/dt}{\sqrt{g_{\alpha\beta}\dot{x}^{\alpha}\dot{x}^{\beta}}%
}\right) _{t=t_{0}}  &  =\frac{1}{c},\\
A_{0}  &  =0,\\
\vec{A}(\rho)  &  =-\frac{B}{2}\rho\hat{\varphi},\\
\vec{v}  &  =\pm v_{0}\hat{\varphi},
\end{align}
where $B$ denotes the modulus of the magnetic field $\vec{\nabla}\times\vec
{A}=\vec{B}=-B\,\hat{z}$, $\rho_{0}$ is the radius of the circular trajectory,
and $v_{0}$ is the constant norm of the velocity of the muons (here $\vec
{v}=\pm v_{0}\hat{\varphi}$ corresponds to the velocity of the muons $\mu
^{\pm}$, respectivelly).

Considering a first order expansion in the dimensionless argument of the exponential function from Eq.(\ref{propertime1}), and using definition (\ref{lambda1}), we approximate the proper time (\ref{propertime1}) as
\end{subequations}
\begin{equation}
\tau=\gamma^{-1}t-\lambda\,\frac{\gamma^{-1}}{2}\int_{0}^{t}\left(  \int
_{0}^{u}\vec{A}\cdot\vec{v}\,ds\right)  du,
\end{equation}
where $\gamma=\frac{1}{\sqrt{1-v_{0}^{2}/c^{2}}}$ is the Lorentz factor.
Solving this equation for the above conditions and writing $v_{0}%
=c\sqrt{1-\gamma^{-2}}$, we obtain
\begin{equation}
\tau(\mu^{\pm})=\gamma^{-1}t(\mu^{\pm})\pm\lambda\frac{\gamma^{-1}}{8}%
c\rho_{0}B\sqrt{1-\gamma^{-2}}\,\left[  t(\mu^{\pm})\right]  ^{2}.
\end{equation}
In the above expression, $\tau(\mu^{\pm})$ is the decay time of the muons
$\mu^{\pm}$ at rest (i.e., in its proper frame) and $t$ is their decay time in
the laboratory frame. By solving the above equation for $t$, the dilated
lifetime $t^{W}$ of the muons, due to Weyl geometry, will be given by
\begin{equation}
t^{W}(\mu^{\pm})=t^{\text{SR}}(\mu^{\pm})\mp\frac{\lambda}{8}c\rho_{0}%
B\sqrt{1-\gamma^{-2}}\left[  t^{\text{SR}}(\mu^{\pm})\right]  ^{2},
\label{tw1}%
\end{equation}
where we have set $t^{\text{SR}}(\mu^{\pm})\doteq\gamma\tau(\mu^{\pm})$ to
indicate the dilated lifetime of the muons, due only to special relativistic
effects, i.e., to the first clock effect. We now see that the magnitude of the
effect increases with the intensity of the magnetic field, the radius of the
trajectory, the value of the Lorentz factor and (quadratically) with the
dilated lifetime.

The parameters of the experiment are
\begin{subequations}
\label{data1}%
\begin{align}
B  &  =1.472\,\text{T},\\
\rho_{0}  &  =7.00\,\text{m},\\
\gamma &  \approx29.327. \label{gamma1}%
\end{align}
Let us note that the above Lorentz factor used in CERN \cite{Bailey:1977de}
and in the E821 experiment \cite{Bennett:2006fi} is called \textit{magic}
$\gamma$, and it is the one that removes the contribution of the stabilizing
quadrupole electrostatic field from the muon's relation between the angular
frequency and the electromagnetic field, named Thomas-Bargmann-Michel-Telegdi
equation \cite{Jegerlehner:2008zza}. The values of the dilated lifetime
$t^{\text{EXP}}(\mu^{\pm})$ of the muons, obtained in the experiment mentioned
above are slightly different from the theoretical values $t^{\text{SR}}%
(\mu^{\pm})$ predicted by special relativity with a precision of $\sim0.1\%$
\cite{Bailey:1977de}:
\end{subequations}
\begin{subequations}
\begin{align}
t^{\text{EXP}}(\mu^{-})  &  \approx64.368\,\mu\text{s},\\
t^{\text{SR}}(\mu^{-})  &  \approx64.461\,\mu\text{s},
\end{align}
and
\end{subequations}
\begin{subequations}
\begin{align}
t^{\text{EXP}}(\mu^{+})  &  \approx64.419\,\mu\text{s},\\
t^{\text{SR}}(\mu^{+})  &  \approx64.431\,\mu\text{s}.
\end{align}
For $\mu^{+}$ and $\mu^{-}$ there were two and four runs respectively, and
what we have presented here is the average of the results. For the calculation
of $t^{\text{SR}}(\mu^{\pm})=\gamma\tau(\mu^{\pm})$, we considered $\tau
(\mu^{\pm})$ as given as in ref. \cite{Bailey:1977de}, which had already been
measured \ (see \cite{Balandin:1975fe,Nordberg:1967zz}).

It is important to remark that the experimentally measured decay times are
distorted by statistical error and systematic effects (described in
\cite{Bailey:1977de,Farley:1979yb}), such as the loss of muons from the
trapping region before decay, variation of the decay electron detection
efficiency (gain effects) and protons that may be stored in the ring, which
contribute to background in the detectors (this effect is significant just for
the $\mu^{+}$ decay time).

Let us now look at the difference between the decay times of the different
muons $\Delta t=t^{\text{SR}}-t^{\text{EXP}}$
\end{subequations}
\begin{align}
\Delta t(\mu^{+})  &  =0.012\,\mu\text{s},\\
\Delta t(\mu^{-})  &  =0.093\,\mu\text{s},
\end{align}

Statistical error and systematic contributions distort the detected lifetime
of the particles. The negative muon $\mu^{-}$ seems to have a smaller detected
decay time than $\mu^{+}$ does, which may be caused by a major contribution of
systematic effects. But since we are also considering the possibility of the
second clock effect, it seems reasonable to suppose that it contributes to the
difference between $\mu^{-}$ and $\mu^{+}$. In this way, it seems plausible to
consider that the non-metricity is diminishing the lifetime of $\mu^{-}$ and
increasing of $\mu^{+}$ (which is compensated by the statistical error and
systematic effects). This situation occurs if $\lambda<0$. Assuming this
scenario we can use the expression (\ref{tw1}) for $\mu^{-}$ to find a constraint.

Let us separate the detected decay time as follows
\begin{equation}
t^{\text{EXP}}=t^{\text{THE}}-t^{\text{ERR}}, \label{def1}%
\end{equation}
where $t^{\text{THE}}$ is the theoretically predicted decay time, which in our
case is given by $t^{\text{W}}$ in Eq.(\ref{tw1}); and $t^{\text{ERR}}\geq0$
is the contribution due to statistical error plus systematic effects of the
experiment for each type of muon. Thus, we have the following inequality:

\begin{equation}
|\lambda|=8\frac{\Delta t(\mu^{-})-t^{\text{ERR}}}{c\rho_{0}B\sqrt
{1-\gamma^{-2}}\left[  t^{\text{SR}}(\mu^{-})\right]  ^{2}}\leq8\frac{\Delta
t(\mu^{-})}{c\rho_{0}B\sqrt{1-\gamma^{-2}}\left[  t^{\text{SR}}(\mu
^{-})\right]  ^{2}}\,,
\end{equation}

which implies, using the above data (\ref{data1}),\footnote{We set
$c\approx2.99\times10^{8}\,\text{m}/\text{s}$.} the following constraint in the CGS system of units:
\begin{equation}
|\lambda|\leq5.81\times10^{-16}\,\text{G}^{-1}\text{cm}^{-2}. \label{bound1}%
\end{equation}
The same order of magnitude would have been found if we had considered
$\mu^{+}$ for the analysis. Therefore, if there exists a second clock effect
caused by Perlick's proper time, then the Weyl parameter $\lambda$ should not
exceed the constraint, i.e., we can estimate an upper bound given by
\begin{equation}
|\lambda|\lessapprox\mathcal{O}(10^{-16})\,\text{G}^{-1}\text{cm}^{-2}.
\label{bound2}%
\end{equation}

In this experiment we have all the ingredients for setting an unprecedented
constraint on the possible existence of the second clock effect. Let us note
that we have followed a conservative approach in the sense that if the Weyl
parameter $\lambda$ does not surpass the order of magnitude (\ref{bound2}),
then the existence of the effect cannot be ruled out.


\subsection{Phenomenological possibilities} 
Phenomenologically we can draw some
scenarios we could expect to appear in future experiments. For instance,
suppose that $\lambda$ is a parameter that depends on the electric charge of
the particle that probes the space-time, i.e., there exists a $\lambda_{+}$
and a $\lambda_{-}$.

$1)$ \ The scenario in which $\lambda_{+}=\lambda_{-}=\lambda$ is suitable for
this experiment, since Eq.(\ref{tw1}) implies that the different types of
muons, i.e., with different charges $\mu^{+}$ and $\mu^{-}$, when accelerated
by a magnetic field will present different decay times. In other words, if
systematic effects equivalently affect the experiment for both types of
particles, then $\Delta t(\mu^{-})$ should be different than $\Delta t(\mu
^{+})$ (assuming the existence of the second clock effect). Moreover, future
experiments measuring dilated lifetimes, would indicate a tendency of increase
in the quantity $|D_{\pm}|$ that we define below:
\begin{equation}
D_{\pm}=\Delta t(\mu^{-})-\Delta t(\mu^{+})
\end{equation}
$2)$ \ If $\lambda_{+}\neq\lambda_{-}$ we should see a different pattern for
$|D_{\pm}|$. For instance, if $\lambda_{+}=-\lambda_{-}$, then we should
observe a decrease in $|D_{\pm}|$.

If we define $\Delta t^{\text{ERR}}\doteq t^{\text{ERR}}(\mu^{-}%
)-t^{\text{ERR}}(\mu^{+})$, \ then we can get rid of these effects by
considering the following analysis for the two different scenarios cited
above:
\begin{equation}
\hspace*{-4cm}1)\ \lambda_{-}=\lambda_{+}=\lambda, \text{which leads
to}\nonumber
\end{equation}
\begingroup\makeatletter

\check@mathfonts%
\begin{equation}
D_{\pm}=\Delta t^{\text{ERR}} -\frac{\lambda}{8} c\rho_{0}B\sqrt{1-\gamma
^{-2}}\left[  \left(  t^{\text{SR}}(\mu^{-})\right)  ^{2}+\left(
t^{\text{SR}}(\mu^{+})\right)  ^{2}\right]  ,
\end{equation}
\endgroup
\begin{equation}
\hspace*{-4cm} 2) \ \lambda_{-}=-\lambda_{+}=\lambda, \text{which
gives}\nonumber
\end{equation}
\begingroup\makeatletter

\check@mathfonts%
\begin{equation}
D_{\pm}=\Delta t^{\text{ERR}} -\frac{\lambda}{8} c\rho_{0}B\sqrt{1-\gamma
^{-2}}\left[  \left(  t^{\text{SR}}(\mu^{-})\right)  ^{2}-\left(
t^{\text{SR}}(\mu^{+})\right)  ^{2}\right]  .
\end{equation}
\endgroup
We expect $\Delta t^{\text{ERR}}$ to be a small quantity that should reduce
even more with the improvement of the experiments. In any case, we can set a
constraint on the parameter $\lambda$ for the two scenarios
\begingroup\makeatletter

\check@mathfonts%
\begin{equation}
|\lambda_{1,2}|\lessapprox8\frac{|\Delta t(\mu^{-})-\Delta t(\mu^{+})|}%
{c\rho_{0}B\sqrt{1-\gamma^{-2}}}\left|  \left(  t^{\text{SR}}(\mu^{-})\right)
^{2}\pm\left(  t^{\text{SR}}(\mu^{+})\right)  ^{2}\right|  ^{-1}.
\end{equation}
\endgroup
Using the above data (\ref{data1}), we find
\begin{align}
|\lambda_{1}|\lessapprox2.53\times10^{-16}\, \text{G}^{-1}\text{cm}%
^{-2},\label{c1}\\
|\lambda_{2}|\lessapprox5.44\times10^{-13}\, \text{G}^{-1}\text{cm}^{-2}.
\end{align}
Since in the second scenario the contributions from the deformed proper time
for $\mu^{+}$ and $\mu^{-}$ compensate each other, the constraint is less
stringent than the one coming from the first scenario, which is of the same
order of magnitude as (\ref{bound2}).

\section{Final remarks}\label{sec-conc}
By analyzing data obtained in an experiment carried
out in CERN which measured the dilated lifetime of muons in a magnetic field
we were able to set constraints on the existence of the second clock effect.
More specifically, in terms of the parameter $\lambda$\ we found that
\begin{equation}
|\lambda|\lessapprox\mathcal{O}(10^{-16})\,\text{G}^{-1}\text{cm}^{-2}.
\end{equation}
Within the limits of experimental accuracy, we are allowed to consider the
second clock effect as being responsible for an extra contribution to the
particle's dilated lifetime, in addition to the one coming from the first
clock effect.

Finally, we would like to call attention for some phenomenological
possibilities that could be addressed in future experiments, for instance the
one to be carried out in Fermilab
\cite{Grange:2015fou,Hertzog:2015jru,Venanzoni:2014ixa}, which is expected to
be operating in the near future. We also mention here the J-PARC experiment
\cite{Saito:2012zz,Sato:2017sdn}, which although uses a different technique
for the measurements of the anomalous magnetic moment of the muon, could also
be of interest for testing the second clock effect.
\newline

\section*{Acknowledgements}
The authors thank Coordena\c c\~ao de Aperfei\c
coamento de Pessoal de N\'ivel Superior (CAPES) and Conselho Nacional de
Desenvolvimento Cient\'ifico e Tecnol\'ogico (CNPq) for financial support.

\end{document}